# Piezoelectric Transition in a Nonpyroelectric Gyroidal Metal-Organic Framework

Shunsuke Kitou[1,*], Hajime Ishikawa[2,*], Yusuke Tokunaga[1], Masato Ueno[1], Hiroshi Sawa[3], Yuiga Nakamura[4], Yuto Kinoshita[2], Tatsuya Miyamoto[5], Hiroshi Okamoto[1], Koichi Kindo[2], and Taka-hisa Arima[1,6]

[1]*Department of Advanced Materials Science, The University of Tokyo, Kashiwa 277-8561, Japan*

[2]*Institute for Solid State Physics, The University of Tokyo, Kashiwa 277-8581, Japan*

[3]*Department of Applied Physics, Nagoya University, Nagoya 464-8603, Japan*

[4]*Japan Synchrotron Radiation Research Institute (JASRI), SPring-8, Hyogo 679-5198, Japan*

[5]*Department of Engineering, Nagoya Institute of Technology, Nagoya 466-8555, Japan*

[6]*RIKEN Center for Emergent Matter Science (CEMS), Wako 351-0198, Japan*

*kitou@edu.k.u-tokyo.ac.jp (SK), hishikawa@issp.u-tokyo.ac.jp (HI)

**Abstract:**

Among the thirty-two crystallographic point groups, 432 is the only one that lacks an inversion center but does not exhibit piezoelectricity. A gyroidal structure belongs to point group 432 and shows characteristic physical properties attributed to its distinctive strong isotropic network. Here, we investigate a gyroidal cobalt oxalate metal-organic framework (MOF) with disordered orientations of $SO_4$ tetrahedra. Synchrotron X-ray diffraction experiments using a single crystal reveal a cubic-to-cubic structural phase transition at $T_S$ = 120 K. This transition involves a change in the point group from nonpiezoelectric 432 to piezoelectric 23. The symmetry change arises from the ordering of distorted SO4 molecules, leading to a three-dimensional helical arrangement of electric dipole moments. Furthermore, pyroelectric current measurements using polycrystalline pellet samples reveal that electric polarization emerges below $T_S$ depending on the magnitude of the pelletizing pressure, demonstrating piezoelectricity. The gyroidal MOF offers an opportunity to explore unique dielectric properties induced by the helical ordering of molecules and structural flexibility.

**INTRODUCTION**

Crystal properties are determined by symmetry. Crystals fall into thirty-two point groups (Figure 1a), with eleven centrosymmetric and twenty-one non-centrosymmetric groups. Among the latter, twenty are piezoelectric, generating polarization under stress. Ten pyroelectric groups can exhibit spontaneous polarization without stress (Figure 1b). If this polarization can be reversed by an external electric field, the material is ferroelectric. Point group 432 lacks an inversion center but is not piezoelectric. If a material in this group transitions to a piezoelectric phase due to temperature change, its polarization and domains can be controlled by anisotropic stress and an electric field during cooling. However, the response of materials that undergo a phase transition from nonpiezoelectric 432 (Figure 1c) to a piezoelectric one (Figure 1d) has not been extensively explored.

Metal-organic framework (MOF) materials provide diverse structures and functionalities[1-7] thanks to their high flexibility and strong response to environmental changes, resulting in a variety of applications, such as gas adsorption,[8,9] catalysis,[10,11] and sensors.[12-14] In the search for novel piezoelectric properties, we focus on MOFs with point group 432. $[(Me_2NH_2)_3(SO_4)]_2[M_2(ox)_3]$ (Me = $CH_3$, $M$ = transition metal, and ox = $C_2O_4$) is a MOF with a gyroidal cubic crystal structure,[15-17] as shown in Figure 2a. The gyroidal structure features a three-dimensional chiral network that includes both three-fold rotation and four-fold screw axes, and it belongs to the space group $I4_132$. This structure is recognized by various names, such as hyperoctagon lattice, $K_4$ crystal, (10,3)-a network, and diamond twin. Alongside honeycomb and diamond structures, the gyroid is one of the three structures mathematically recognized for its strong isotropic properties.[18] Gyroidal MOFs are expected to exhibit unique physical properties due to their distinctive geometric network and electronic structure.[19-24]

Here, we study a gyroidal cobalt oxalate (gy-Co-ox), $[(Me_2NH_2)_3(SO_4)]_2[Co_2(ox)_3]$, whose magnetic properties, arising from the Co ions, have attracted attention in the field of quantum magnetism.[17,25] At room temperature, gy-Co-ox has a nonpiezoelectric nature characterized by the point group 432. Co and S occupy the Wyckoff positions of 8*a* and 8*b* of $I4_132$, respectively, with the common site symmetry of .32. They form a gyroidal lattice with opposite chirality (Figure 2b). Within a primitive unit cell, there are four sub-lattices for the Co and S sites. The direction of the local three-fold rotation axis is different at each sublattice: the three-fold axis is along the [111] axis and its equivalents. The $SO_4$ tetrahedron, surrounded by six $Me_2NH_2$ molecules, exhibits orientational disorder,[15] as shown in Figure 2c. The $SO_4$ tetrahedron shrinks along the three-fold axis and can point either parallel or antiparallel to the axis. This corresponds to the Ising-like degrees of freedom of the electric dipole moment. If the SO$_4$ molecules are to orderly arrange, a reduction in symmetry is expected.

In this study, we observe a cubic-to-cubic phase transition at 120 K by single-crystal X-ray diffraction (XRD) experiments of gy-Co-ox. This phase transition is induced by the ordering of distorted SO$_4$ molecules, resulting in the formation of a three-dimensional helical arrangement of electric dipole moments on the SO$_4$ molecules. Furthermore, through pyroelectric current measurements using pellet samples of gy-Co-ox, electric polarizations associated with a nonpyroelectric piezoelectric transition are observed.

**RESULTS AND DISCUSSION**

The inset of Figure 3a shows the temperature dependence of the magnetic susceptibility $\chi$ of gy-Co-ox polycrystals. As reported in the previous study,[17] gy-Co-ox exhibits magnetic transitions arising from the Co ions at $T_H$ = 18 K and $T_L$ = 10 K, where $T_H$ and $T_L$ indicate the high and low magnetic-transition temperatures, respectively. Figure 3a shows the temperature dependence of specific heat divided by temperature, $C_p/T$. A peak is observed at $T_S$ = 120 K in $C_p/T$, whereas no anomaly is detected at $T_S$ in $\chi$. $T_S$ corresponds to the structural phase transition temperature, as will be shown later. The temperature dependence of the dielectric constant also shows a broad peak at $T_S$ = 120 K, as shown in Figure 3b.

Figure 2c shows the results of the single-crystal structure analysis of gy-Co-ox at 140 K, which is consistent with the structure reported in the previous study.[15] The orientational disorder is observed in the SO$_4$ molecule, which is surrounded by six Me$_2$NH$_2$ molecules. The SO$_4$ tetrahedron is slightly distorted due to hydrogen bonding (see Figure S2 in SI). Similar distortions of SO$_4$ tetrahedra caused by hydrogen bonding have also been reported in well-known crystals, such as alum KAl(SO$_4$)$_2 \cdot$12H$_2$O[26] and copper sulfate CuSO$_4 \cdot$5H$_2$O.[27] No merohedral domain corresponding to space inversion (-$a$,-$b$,-$c$) is found in this single crystal of 100 × 100 × 90 µm$^3$ (see Table S1 in SI). On cooling, we observe the appearance of 2 0 0 peak in single-crystal XRD data below $T_S$ = 120 K (Figures 3c and 3d). The 2 0 0 peak does not satisfy the reflection condition for the 4$_1$ screw axis along [100] ($H$ 0 0: $H$ = 4$n$) of the space group $I4_132$. The reflection condition derived from the $I$-lattice ($H$ $K$ $L$: $H$ + $K$ + $L$ = even) is preserved, and no Bragg peak splitting is observed below $T_S$ (Figure S1 in SI). These results indicate the space group $I2_13$ in the low-temperature phase. It is noted that, among the maximal non-isomorphic subgroups of $I4_132$, $I2_13$ is the only space group that preserves the $I$-lattice cubic structure.

The structural analysis was performed assuming the space group $I2_13$ using single-crystal XRD data at 100 K, resulting in convergence of all structural parameters. We also observe the polarization dependence of second harmonic generation signal, which is consistent with the symmetry of point group 23 (Figure S10 in SI). Generally, structural phase transitions are often accompanied by lattice distortions. There are only a few reports of cubic-to-cubic phase transitions. $C_{60}$ fullerene undergoes a cubic-to-cubic phase transition with decreasing temperature,[28,29] which is attribute to the rotational ordering of $C_{60}$ molecules. This transition occurs from the space group $Fm$-$3m$ to $Pa$-$3$, both of which belong to the achiral centrosymmetric point groups ($m$-$3m$ and $m$-$3$). As the $F$-lattice changes to a $P$-lattice, the primitive cell becomes four times larger. β-pyrochlore $CsW_2O_6$[30] and lacunar spinel $GaNb_4Se_8$[31,32] also exhibit cubic-to-cubic transitions with decreasing temperature, attributed to the orbital and charge ordering of transition metals. The former is a transition from the space group $Fd$-$3m$ to $P2_13$, and the latter is a transition from $F$-$43m$ to $P2_13$. In both systems, the phase transition involves a change in point group from achiral ($m$-$3m$ and -$43m$ in $CsW_2O_6$ and $GaNb_4Se_8$) to chiral (23 in both), accompanied by a fourfold increase in the size of the primitive unit cell. On the other hand, in the case of gy-Co-ox, the point groups in the high- and low-temperature phases are 432 and 23, respectively. Therefore, gy-Co-ox not only maintains the $I$-lattice cubic structure during the phase transition but also retains chirality in both the high- and low-temperature phases. Furthermore, the cubic-to-cubic transition in $C_{60}$, $CsW_2O_6$, and $GaNb_4Se_8$ proceeds via multiple intermediate maximal subgroups ($Fm$-$3m$ → $Fm$-$3$ → $Pa$-$3$, $Fd$-$3m$ → $F$-$43m$ → $F23$ → $P2_13$, and $F$-$43m$ → $F23$ → $P2_13$, respectively). In contrast, in the case of gy-Co-ox, $I2_13$ is a maximal subgroup of $I4_132$, meaning that the crystallographic symmetry reduction occurs in a single step. To the best of our knowledge, there have been no reports of a cubic-to-cubic transition involving a change in chiral point group from nonpiezoelectric 432 to piezoelectric 23 without altering the primitive cell size.

Figure 4a shows the crystal structure at 100 K in the low-temperature phase. While the gyroid network of Co ions remains hardly changed (Figure S4 in SI), the orientations of $SO_4$ molecules are regularly arranged in the low-temperature phase. It is noted that the ordering is incomplete, with approximately one-third of the $SO_4$ molecules remaining disordered (Figure S3 in SI). Since the $SO_4$ molecule possesses a local electric dipole moment, applying an external electric field may help reduce the degree of residual disorder. Additionally, as the phase transition in this system involves the disorder-order transition of the $SO_4$ molecules, the cooling rate may also influence the residual disorder. The Ising model for the $SO_4$ ordering predicts that the entropy released per $SO_4$ molecule is $R\ln2 \sim 5.46$ J/(K mol). The estimated value of 2.2 J/(K mol) per $SO_4$ molecule from the temperature dependence of $C_p/T$ (Figure 3a) seems to be compatible with the remaining disorder.

In the high-temperature disordered phase, the shortest distance between the basal O2 atoms of SO$_4$ tetrahedron and the H atoms of NH$_2$ is 2.062(13) Å (Figure 2c). The ordering of the SO$_4$ molecules leads to the formation of hydrogen bonds with the surrounding three Me$_2$NH$_2$ molecules, as evidenced by the O2–H distances of 1.938(3) Å (Figure 4a), which are shorter than those in the high-temperature phase. The SO$_4$ molecules form an antiferroic arrangement via hydrogen bonding, with the apical O1 atoms in adjacent SO$_4$ molecules pointing in different <111> directions. Here, the SO$_4$ ordering leads to the emergence of a static electric dipole moment, as shown in Figure 4a. Neighboring electric dipole moments, $\bm{p}_i$ and $\bm{p}_j$, along the local threefold axis are arranged in the antiferroic manner ($\bm{p}_i \cdot \bm{p}_j < 0$) through the dipole-dipole interaction. As a result, the *I*-lattice cubic structure is preserved during the phase transition (see Figure S5 in SI). Furthermore, these electric dipole moments are arranged helically along the <111> direction in the crystal (Figure 4b). In this case, since the site symmetry at the S site changes from .32 to .3. by the phase transition, the two-fold rotation axes along <110> disappear. As a result, there can be merohedral domains with opposite dipole moments. The structural analysis identified these domains to exist approximately in a 1:1 ratio within the crystal (see Table S2 in SI). Although the ordering of SO$_4$ molecules and the hydrogen bond network structure play important roles in the phase transition of this system, the underlying mechanism remains controversial. Thermodynamic and kinetic investigations based on hydrogen bond dynamics are anticipated in the future.

Due to the change in point group symmetry from 432 to 23 and the emergence of static local electric dipoles associated with the phase transition in gy-Co-ox, a piezoelectric response is expected in the low-temperature phase. We performed pyroelectric current measurements using gy-Co-ox polycrystals formed into thin cylindrical pellets under a pressure of 10 MPa. Here, pressure was applied only during the pellet formation. While no pressure was applied during the pyroelectric current measurement, residual stress could not be completely removed, as described later. When using a polycrystal, after poling under anisotropic stress and electric field, the piezoelectric response should be observed regardless of the direction of the applied stress. Figures 5a and 5b show the temperature dependence of the pyroelectric current and electric polarization measured at zero electric field, after cooling under different poling electric fields $E_{\text{pol}}$ (see Figure S7 in SI for details of pyrocurrent measurements). The clear phase transitions are observed at $T_S$ = 120 K, and these signals increase with increasing $E_{\text{pol}}$. It was confirmed that the observed peaks in the pyroelectric current were not an artifact due to thermally stimulated current[33] but were instead attributable to intrinsic electric polarization (see Figure S9 in SI).

It should be noted that since the low-temperature phase of gy-Co-ox belongs to the nonpyroelectric piezoelectric point group 23, macroscopic electric polarization may not appear unless external stress is applied. We consider that the residual anisotropic stress in the pellet, caused by the softness of the MOF-derived structure, might allow electric polarization to emerge. To verify this hypothesis, we examine the pelletizing pressure dependence of the piezoelectric response. Pyroelectric current measurements on gy-Co-ox polycrystals, pelletized under a lower pressure of 0.5 MPa, result in smaller electric polarization (Figures 5c and 5d). When the pelletizing pressure is increased up to 6 MPa (Figures 5e and 5f), the magnitude of the electric polarization increases monotonically. This suggests the presence of residual stress exists in the pellet samples. The observed electric polarization, with a magnitude of approximately 30 $\mu C/m^2$ at 10 MPa, is 3 to 4 orders of magnitude smaller than the previously reported polarization values in MOF systems[34-38] (see Table S3 in SI). This cubic MOF exhibits electric dipole moments aligned along four equivalent <111> directions. In the absence of uniaxial stress, these dipoles cancel each other out, leading to a small piezoelectric response. The low response may also be due to measurements under residual stress rather than uniaxial stress, the use of polycrystalline pellets instead of a single crystal, and incomplete domain alignment (Figure S8 in SI).

In our polarization measurements using a polycrystalline pellet, grains of positive and negative piezoelectric coefficient are nearly equally distributed, resulting in negligible macroscopic electric polarization in the absence of a poling electric field (black lines in Figures 5b and 5d). On the other hand, after the pellet is cooled across the transition temperature in a poling electric field $E_{pol}$ under possible residual stress induced by pelletizing, the piezoelectric domains are roughly aligned in the 23 point group (see Figure S7 in SI), leading to the appearance of macroscopic electric polarization. Furthermore, the electric polarization is reversed by changing the sign of $E_{pol}$. It is worth noting that different merohedral domains should exhibit the opposite signs of piezoelectricity. This change indicates that the merohedral domains induced by the stress can be controlled by $E_{pol}$. Our results are perfectly consistent with the electric polarization response corresponding to the change in point group from nonpiezoelectric 432 to piezoelectric 23.

Although several recent papers have reported anion ordering in a cubic system of 432 point group, the anion molecules are ordered above room temperature, leading to tetragonal pyroelectric point group 4 at room temperature.[34-36] In contrast, the MOF system in the present study (i) undergoes a disorder-order transition of anion molecules below room temperature, and (ii) exhibits a change in cubic point group symmetry from nonpiezoelectric 432 to piezoelectric 23 without any lattice distortion, which has never been reported. Materials exhibiting such symmetry changes hold significant potential for advanced applications in functional materials science. This system demonstrates a remarkable phase transition from a nonpiezoelectric, nonpyroelectric structure to a piezoelectric, nonpyroelectric structure. This distinctive property can be harnessed to design robust, low-power memory devices that exploit the system's bistability, which is controllable via mechanical stress and temperature. Such devices may play a crucial role in the development of beyond-CMOS technologies. Moreover, the preservation of cubic lattice symmetry during the phase transition ensures minimal lattice distortion, a feature that translates to exceptional durability under cyclic operations. This combination of unique functional properties and structural stability could position the material with such symmetry as a promising candidate for next-generation electronic and electromechanical applications.

## CONCLUSIONS

We discover the nonpiezoelectric to nonpyroelectric piezoelectric transition at $T_S$ = 120 K of gy-Co-ox by synchrotron XRD experiments using a single crystal. This MOF system maintains the chiral cubic lattice across the phase transition. The structural transition is characterized by the orientational ordering of trigonally distorted polar $SO_4$ molecules, forming a helical arrangement of the electric dipole moments derived from the gyroidal structure. Pyroelectric current measurements using pellet samples reveal electric polarization, the magnitude of which depends on the pelletizing pressure below $T_S$. Moreover, gy-Co-ox possesses not only a unique helical arrangement of the electric dipoles in $SO_4$ but also a spin degree of freedom at the Co site. The Co ions have the same site symmetry as the S and exhibit a variety of magnetic phases at low temperatures.[17] Therefore, exploring the properties emerging from the combination of the two different degrees of freedom arranged on the gyroidal lattice would also be of interest. This system, with its unique features such as the symmetry change between two chiral cubic point groups, the multiple degrees of freedom, and the inherent flexibility of MOFs, opens up the search of novel responses to various external fields such as electric and magnetic fields, stress, and light.

## MATERIALS AND METHODS

**Sample Preparation.** The ample was prepared by a solvothermal method. In a typical synthesis, $CoSO_4·7H_2O$ (0.281g, 1 mmol) and $H_2C_2O_4$ (0.180 g, 2 mmol) were put in a glass vial with 2.5 mL of N,N-dimethylformamide (DMF). The mixture was homogenized by sonication and transferred into a Teflon-lined stainless-steel autoclave. The autoclave was heated at 180℃ for a few days. The pink polycrystalline sample including small single crystals of ~ 100 μm size was obtained. The sold was collected by decantation and washed by DMF a few times. The sample was vacuum dried and stored in an argon filled glovebox to prevent degradation by moisture.

**XRD Experiments.** XRD experiments were performed on BL02B1 at a synchrotron facility SPring-8, Japan.[39] An $N_2$-gas-blowing device was employed to cool the crystals to 100 K. A two-dimensional detector CdTe PILATUS was used to record the diffraction pattern. The X-ray wavelength was $\lambda = 0.30960$ Å. The intensities of Bragg reflections of the interplane distance $d > 0.5$ Å were collected by CrysAlis program.[40] Intensities of equivalent reflections were averaged and the structural parameters were refined by using Jana2006.[41] Crystal structures are visualized by using VESTA.[42]

**Magnetic Susceptibility.** Magnetization of the polycrystal sample was measured by a SQUID magnetometer in a temperature range between 1.8 and 300 K at 1 T (MPMS-XL, Quantum Design).

**Specific Heat.** Specific heat of the polycrystalline sample was measured by the relaxation method using a commercial apparatus (PPMS, Quantum Design) between 2 and 200 K. The polycrystalline powder was pelletized for the measurements.

**Dielectric Constant.** Dielectric constant on the pelletized sample was measured by a conventional two-terminal method using an LCR meter (E4980A, Agilent).

**Electric Polarization.** Electric polarization was obtained by integrating the pyroelectric current measured by using an electrometer (6517A, Keithley), while heating the pellet at a rate of 6 K/min. The polycrystalline powder was pelletized for the measurement.

## SUPPORTING INFORMATION

Additional experimental data of XRD, electric polarization, second harmonic generation.

## ACKNOWLEDGMENT

This work was supported by a Grant-in-Aid for Scientific Re-search (No. 21H04988, 22K14010, 23H01120, 24H01644, and 24H01650) from JSPS. The synchrotron radiation experiments were performed at SPring-8 with the approval of the Japan Synchrotron Radiation Research Institute (JASRI) (Proposal No. 2023B0304).

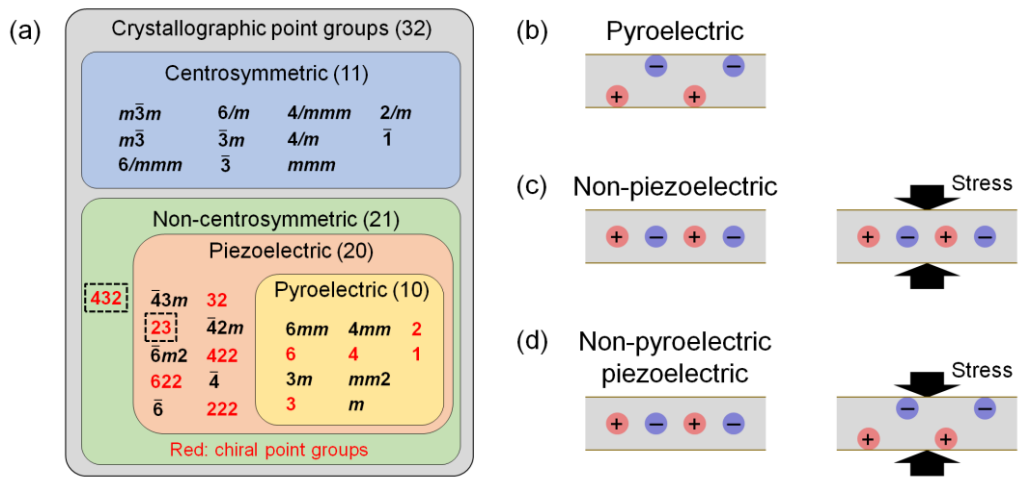

Figure 1. (a) Classification of the thirty-two crystallographic point groups. Schematic representation of macroscopic electric polarization in (b) pyroelectric, (c) nonpiezoelectric, and (d) nonpyroelectric piezoelectric point groups.

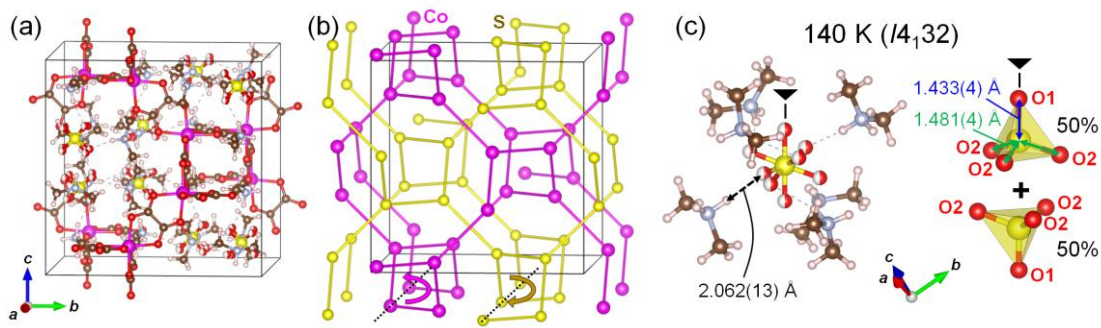

Figure 2. (a) Crystal structure of $[(Me_2NH_2)_3(SO_4)]_2[Co_2(ox)_3]$ at 140 K in the high-temperature phase. Co and S atoms are represented by pink and yellow, respectively. (b) Gyroidal network formed by Co and S atoms, exhibiting chirality opposite to each other. (c) Local structure around an $SO_4$ molecule, surrounded by six $Me_2NH_2$ molecules. Solid triangles indicate three-fold rotation axes along [111].

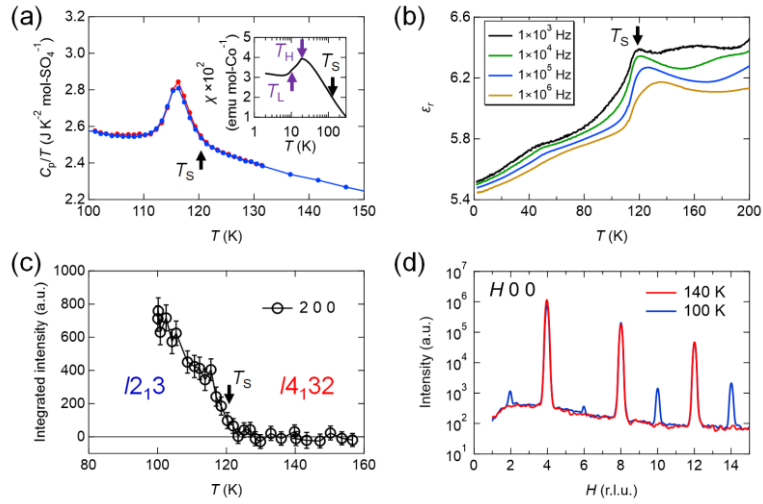

Figure 3. (a) Temperature dependence of specific heat divided by temperature, $C_p/T$. Blue and red dots indicate the data in the cooling and heating processes, respectively. The inset shows the temperature dependence of magnetic susceptibility $\chi$ for $\mu_0 H = 1$T. (b) Temperature dependence of dielectric constant measured at several frequencies. (c) Temperature dependence of the integrated intensity of the 2 0 0 reflection. (d) One-dimensional semilogarithmic plots of the XRD profile along the $H$ 0 0 line.

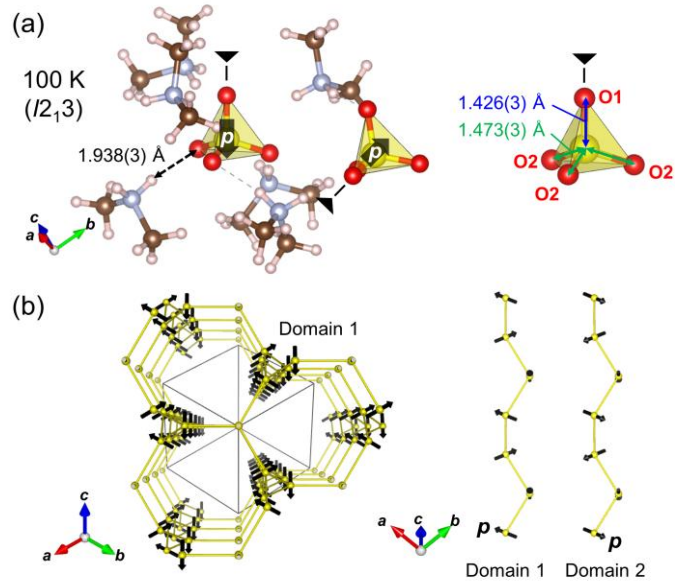

Figure 4. (a) Local structure around $SO_4$ molecules at 100 K in the low-temperature phase. The black arrows labeled ***p*** indicate the electric dipole moments of trigoally distorted $SO_4$ molecules. (b) Perspective projection of helical arrangement of electric dipole moments on S atoms viewed along the [111] axis. There are two types of merohedral domains depending on the orientation of the dipole moments.

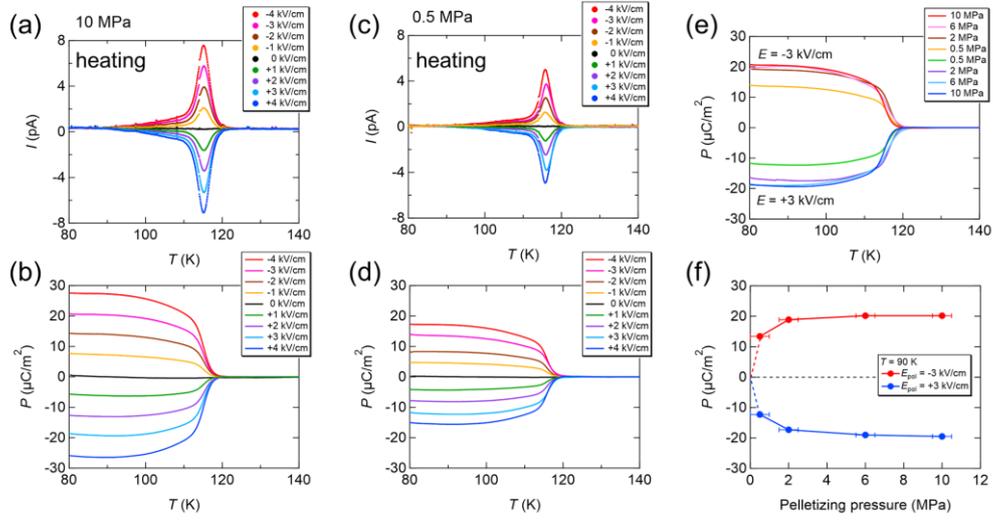

Figure 5. Temperature dependence of (a,c) pyroelectric current and (b,d) electric polarization of polycrystals pelletized under pressures of 10 and 0.5 MPa, respectively, measured at zero electric field after cooling in various poling electric fields $E_{pol}$. (e) Temperature dependence of electric polarization of polycrystals pelletized under various pressures. (f) Pelletizing-pressure dependence of electric polarization at 90 K after cooling in $E_{pol}$ = ±3 kV/cm.

**References**

1. Kitagawa, S.; Kitaura, R.; Noro, S. Functional Porous Coordination Polymers. *Angew. Chem. Int. Ed.* **2004**, *43*, 2334-2375.
2. Zhang, W.; Xiong, R.-G. Ferroelectric Metal–Organic Frameworks. *Chem. Rev.* **2012**, *112*, 1163-1195.
3. Stroppa, A.; Barone, P.; Jain, P.; Perez-Mato, J. M.; Picozzi, S. Hybrid Improper Ferroelectricity in a Multiferroic and Magnetoelectric Metal-Organic Framework. *Adv. Mater.* **2013**, *25*, 2284-2290.
4. Furukawa, H.; Cordova, K. E.; O'Keeffe, M.; Yaghi, O. M. The Chemistry and Applications of Metal-Organic Frameworks. *Science* **2013**, *341*, 1230444.
5. Zhou, H.-C.; Kitagawa, S. Metal–Organic Frameworks (MOFs). *Chem. Soc. Rev.* **2014**, *43*, 5415-5418.
6. Ding, M.; Cai, X.; Jiang, H.-L. Improving MOF stability: approaches and applications. *Chem. Sci.* **2019**, *10*, 10209-10230.
7. Freund, R.; Zaremba, O.; Arnauts, G.; Ameloot, R.; Skorupskii, G.; Dincă, M.; Bavykina, A.; Gascon, J.; Ejsmont, A.; Goscianska, J.; Kalmutzki, M.; Lächelt, U.; Ploetz, E.; Diercks, C. S.; Wuttke, S. The Current Status of MOF and COF Applications. *Angew. Chem. Int. Ed.* **2021**, *60*, 23975-24001.
8. Li, J.-R.; Tao, Y.; Yu, Q.; Bu, X.-H.; Sakamoto, H.; Kitagawa, S. Selective Gas Adsorption and Unique Structural Topology of a Highly Stable Guest-Free Zeolite-Type MOF Material with N-rich Chiral Open Channels. *Chem. Eur. J.* **2008**, *14*, 2771-2776.
9. Ghanbari, T.; Abnisa, F.; Daud, W. M. A. W. A review on production of metal organic frameworks (MOF) for $CO_2$ adsorption. *Sci. Total Environ.* **2020**, *707*, 135090.
10. Lee, J. Y.; Farha, O. K.; Roberts, J.; Scheidt, K. A.; Nguyen, S. B. T.; Hupp, J. T. Metal–organic framework materials as catalysts. *Chem. Soc. Rev.* **2009**, *38*, 1450-1459.
11. Wang, Q.; Astruc, D. State of the Art and Prospects in Metal–Organic Framework (MOF)-Based and MOF-Derived Nanocatalysis. *Chem. Rev.* **2020**, *120*, 1438-1511.
12. Kreno, L. E.; Leong, K.; Farha, O. K.; Allendorf, M.; Duyne, R. P. V.; Hupp, J. T. Metal–organic framework materials as chemical sensors. *Chem. Rev.* **2012**, *112*, 1105-1125.
13. Yi, F.-Y.; Chen, D.; Wu, M.-K.; Han, L.; Jiang, H.-L. Chemical Sensors Based on Metal–Organic Frameworks. *Chem. Plus. Chem.* **2016**, *81*, 675-690.

14. Jo, Y.-M.; Jo, Y. K.; Lee, J.-H.; Jang, H. W.; Hwang, I.-S.; Yoo, D. J. MOF-Based Chemiresistive Gas Sensors: Toward New Functionalities. *Adv. Mater.* **2023**, *35*, 2206842.
15. Li, C.-R.; Li, S.-L.; Zhang, X.-M. $D_3$-Symmetric Supramolecular Cation $\{(Me_2NH_2)_6(SO_4)\}^{4+}$ As a New Template for 3D Homochiral (10,3)-a Metal Oxalates. *Cryst. Growth Des.* **2009**, *9*, 1702.
16. Nagarkar, S. S.; Unni, S. M.; Sharma, A.; Kurungot, S.; Ghosh, S. K. Two-in-One: Inherent Anhydrous and Water-Assisted High Proton Conduction in a 3D Metal–Organic Framework. *Angew. Chem. Int. Ed.* **2014**, *53*, 2638-2642.
17. Ishikawa, H.; Imajo, S.; Takeda, H.; Kakegawa, M.; Yamashita, M.; Yamaura, J.; Kindo, K. $J_{eff}$ = 1/2 Hyperoctagon Lattice in Cobalt Oxalate Metal-Organic Framework *Phys. Rev. Lett.* **2024**, *132*, 156702.
18. Sunaga, T. Crystals That Nature Might Miss Creating. *Notices of the AMS* **2008**, *55* 208.
19. Gier, T. E.; Bu, X.; Feng, P.; Stucky, G. D. Synthesis and organization of zeolite-like materials with three-dimensional helical pores. *Nature* **1998**, *395*, 154-157.
20. Zhou, X.-P.; Li, M.; Liu, J. & Li, D. Gyroidal metal–organic frameworks. *J. Am. Chem. Soc.* **2012**, *134*, 67-70.
21. Gao, Q.; Wang, X.-L.; Xu, J.; Bu, X.-H. The First Demonstration of the Gyroid in a Polyoxometalate-Based Open Framework with High Proton Conductivity. *Chem. Eur. J.* **2016**, *22*, 9082-9086
22. Mizuno, A.; Shuku, Y.; Suizu, R.; Matsushita, M. M.; Tsuchiizu, M.; Mañeru, D. R.; Illas, F.; Robert, V.; Awaga, K. Discovery of the $K_4$ Structure Formed by a Triangular π Radical Anion. *J. Am. Chem. Soc.* **2015**, *137*, 7612-7615.
23. Mizuno, A.; Shuku, Y.; Awaga, K. Recent Developments in Molecular Spin Gyroid Research. *Bull. Chem. Soc. Jpn.* **2019**, *92*, 1068-1093.
24. Nakashima, K.; Suizu, R.; Morishita, S.; Tsurumachi, N.; Funahashi, M.; Masu, H.; Ozawa, R.; Nakamura, K.; Awaga, K. Enhanced Circularly Polarized Luminescence by a Homochiral Guest–Host Interaction in Gyroidal MOFs, [Ru(bpy)$_3$] [M$_2$(ox)$_3$] (bpy = 2,2′-Bipyridyl, ox = Oxalate, M = Zn, Mn). *ACS Mater. Au* **2023**, *3*, 201-205.
25. Yamada, M. G.; Fujita, H.; Oshikawa, M. Designing Kitaev Spin Liquids in Metal-Organic Frameworks. *Phys. Rev. Lett.* **2017**, *119*, 057202.
26. Nyburg, S. C.; Steed, J. W.; Aleksovska, S.; Petruševski, V. M. Structure of the alums. I. On the sulfate group disorder in the α-alums. *Acta Cryst.* **2000**, *B56*, 204-209.
27. Varghese, J. N.; Maslen, E. N. Electron density in non-ideal metal complexes. I. Copper sulphate pentahydrate. *Acta Cryst.* **1985**, *B41*, 184-190.

28. Liu, S.; Lu, Y. J.; Kappes, M. M.; Ibers, J. A. The Structure of the $C_{60}$ Molecule: X-ray Crystal Structure Determination of a Twin at 110 K. *Science* **1991**, *254*, 408-410.
29. David, W. I. F.; Ibberson, R. M.; Dennis, T. J. S.; Hare, J. P.; Prassides, K. Structural Phase Transitions in the Fullerene $C_{60}$. *Europhys. Lett.* **1992**, *18*, 219-225.
30. Okamoto, Y.; Amano, H.; Katayama, N.; Sawa, H.; Niki, K.; Mitoka, R.; Harima, H.; Hasegawa, T.; Ogita, N.; Tanaka, Y.; Takigawa, M.; Yokoyama, Y.; Takehana, K.; Imanaka, Y.; Nakamura, Y.; Kishida, H.; Takenaka, K. Regular-triangle trimer and charge orderpreserving the Anderson condition in the pyrochlore structure ofCsW$_2$O$_6$. *Nat. Commun.* **2020**, *11*, 3144.
31. Ishikawa, H.; Yajima, T.; Matsuo, A.; Ihara, Y.; Kindo, K. Nonmagnetic Ground States and a Possible Quadrupolar Phase in 4*d* and 5*d* Lacunar Spinel Selenides Ga*M*$_4$Se$_8$ (*M* = Nb, Ta). *Phys. Rev. Lett.* **2020**, *124*, 227202.
32. Kitou, S.; Gen, M.; Nakamura, Y.; Tokunaga, Y.; Arima, T. Cluster Rearrangement by Chiral Charge Order in Lacunar Spinel GaNb$_4$Se$_8$. *Chem. Mater.* **2024**, *36*, 2993-2999.
33. Terada, N.; Glazkova, Y. S.; Belik, A. A. Differentiation between ferroelectricity and thermally stimulated current in pyrocurrent measurements of multiferroic *M*Mn$_7$O$_{12}$ (*M* = Ca, Sr, Cd, Pb). *Phys. Rev. B* **2016**, *93*, 155127.
34. Yadav, A.; Srivastava, A. K.; Kulkarni, P.; Divya, P.; Steiner, A.; Praveenkumar, B.; Boomishankar, R. Anion-induced ferroelectric polarization in a luminescent metal–organic cage compound. *J. Mater. Chem. C*, **2017**, *5*, 10624-10629.
35. Yadav, A.; Kulkarni, P.; Praveenkumar, B.; Steiner, A.; Boomishankar, R. Hierarchical Frameworks of Metal–Organic Cages with Axial Ferroelectric Anisotropy. *Chem. Eur. J.* **2018**, *24*, 14639-14643.
36. Prajesh, N.; Naphade, D. R.; Yadav, A.; Kushwaha, V.; Praveenkumar, B.; Zaręba, J. K.; Anthopoulos, T. D.; Boomishankar, R. Visualization of domain structure and piezoelectric energy harvesting in a ferroelectric metal–ligand cage. *Chem. Commun.* **2023**, *59*, 2919-2922.
37. Jain, P.; Ramachandran, V.; Clark, R. J.; Zhou, H. D.; Toby, B. H.; Dalal, N. S.; Kroto, H. W.; Cheetham, A. K. Switchable Electric Polarization and Ferroelectric Domains in a Metal-Organic Framework. *npj Quantum Mater.* **2016**, *1*, 16012.
38. Guo, M.; Cai, H.-L.; Xiong, R.-G. Ferroelectric Metal Organic Framework (MOF). *Inorg. Chem. Commun.* **2010**, *13*, 1590-1598.
39. Sugimoto, K.; Ohsumi, H.; Aoyagi, S.; Nishibori, E.; Moriyoshi, C.; Kuroiwa, Y.; Sawa, H.; Takata, M. Extremely High Resolution Single Crystal Diffractometry for Orbital

Resolution using High Energy Synchrotron Radiation at SPring-8. *AIP Conf. Proc.* **2010**, *1234*, 887-890.

40. CrysAlisPro; Agilent Technologies Ltd: Yarnton, **2014**.
41. Petříček, V.; Dušek, M. Palatinus, L. Discontinuous modulation functions and their application for analysis of modulated structures with the computing system JANA2006. *Z. Kristallogr. Cryst. Mater.* **2014**, *229*, 345-352.
42. Momma, K.; Izumi, F. VESTA 3 for three-dimensional visualization of crystal, volumetric and morphology data. *J. Appl. Crystallogr.* **2011**, *44*, 1272-1276.